\documentclass[twocolumn]{article}

\usepackage{PRIMEarxiv}
\usepackage[utf8]{inputenc} 
\usepackage[T1]{fontenc}    
\usepackage{hyperref}       
\usepackage{booktabs}       
\usepackage{microtype}      
\usepackage{amsthm}
\usepackage{bm}
\usepackage{amssymb}
\usepackage{fancyhdr}       
\usepackage{graphicx}       
\graphicspath{{media/}}     
\usepackage{authblk}
\usepackage{subcaption}
\usepackage{makecell}
\usepackage[short,nocomma]{optidef}
\usepackage{tabularx}
\usepackage{siunitx}
\DeclareSIUnit{\hollandseflorijn}{Hfl}
\DeclareSIUnit{\permille}{\text{\textperthousand}}
\usepackage{tikz}
\usepackage{pgfplots}
\usepgfplotslibrary{groupplots}
\pgfplotsset{compat=1.18}
\def\axisdefaultheight{110pt}

\theoremstyle{remark}
\newtheorem{remark}{Remark}

\title{Reinforcement Learning-based Model Predictive Control for Greenhouse Climate Control}

\author[*,1, **]{\textbf{Samuel Mallick}}
\author[1, **]{\textbf{Filippo Airaldi}}
\author[1]{\textbf{Azita Dabiri}}
\author[2]{\textbf{Congcong Sun}}
\author[1]{\textbf{Bart De Schutter}}

\affil[1]{Delft Center for Systems and Control, Delft University of Technology, 2628 CD Delft, The Netherlands}
\affil[2]{Agricultural Biosystems Engineering Group, Wageningen University, 6700 AA Wageningen, The Netherlands}
\affil[*]{Corresponding author, e-Mail: \url{s.h.mallick@tudelft.nl}}
\affil[**]{These authors contributed equally to this work.}

\begin{document}

\twocolumn[
    \begin{@twocolumnfalse}
        \maketitle
        \begin{abstract}
            Greenhouse climate control is concerned with maximizing performance in terms of crop yield and resource efficiency.
One promising approach is model predictive control (MPC), which leverages a model of the system to optimize the control inputs, while enforcing physical constraints.
However, prediction models for greenhouse systems are inherently inaccurate due to the complexity of the real system and the uncertainty in predicted weather profiles.
For model-based control approaches such as MPC, this can degrade performance and lead to constraint violations.
Existing approaches address uncertainty in the prediction model with robust or stochastic MPC methodology; however, these necessarily reduce crop yield due to conservatism and often bear higher computational loads.
In contrast, learning-based control approaches, such as reinforcement learning (RL), can handle uncertainty naturally by leveraging data to improve performance.
This work proposes an MPC-based RL control framework to optimize the climate control performance in the presence of prediction uncertainty.
The approach employs a parametrized MPC scheme that learns directly from data, in an online fashion, the parametrization of the constraints, prediction model, and optimization cost that minimizes constraint violations and maximizes climate control performance.
Simulations show that the approach can learn an MPC controller that significantly outperforms the current state-of-the-art in terms of constraint violations and efficient crop growth.  
        \end{abstract}
        \vspace{24pt}
    \end{@twocolumnfalse}
]

\section{Introduction}
Greenhouse climate control presents a key opportunity to address the growing world population's food production requirements in a changing climate, and the grand societal challenge of efficient energy consumption.
With modern greenhouses equipped with actuation systems such as heating, ventilation, and CO$_2$ injection, effective control approaches can lead to high crop yield in an energy-efficient manner.
However, the control challenge is difficult as the process dynamics are highly nonlinear and complex \cite{vanhentenGreenhouseClimateManagement1994}, and the climate variables, such as temperature and humidity, must be effectively constrained to avoid damage to crops due to, e.g., spread of diseases \cite{chen2020data, xuAdaptiveTwoTimeScale2018,gullino2020integrated}. {\color{black} While traditional methods, e.g., on-off and PID control, have been used for low-level regulation, these strategies are not rooted in optimal control and are in general unable to deliver optimal performance and to systematically handle complex state and/or input constraints \cite{hamzaRobustTSFuzzy2019, lafontModelFreeControlExperimental2013}.}

Model predictive control (MPC) is an optimization-based control methodology that naturally handles multi-input-multi-output systems with state, input, and output constraints \cite{borrelliPredictiveControlLinear2016}.
It has seen huge theoretical success and application in process control, and has been proposed to solve greenhouse control challenges \cite{blascoModelBasedPredictiveControl2007, gruberNonlinearMPCBased2011, montoyaHybridControlledApproachMaintaining2016}.
However, MPC relies heavily on an accurate prediction model, while in a greenhouse prediction model there always exist uncertainties due to, e.g., modeling error and inaccurate weather forecasting.
For model-based control, an incorrect model can lead the controller to drive the system to an undesired point of operation, possibly violating constraints.
Furthermore, as constraints often represent the validity range for the accuracy of the prediction model, violations of the constraints can lead to degraded performance when applied to the real system \cite{xuAdaptiveTwoTimeScale2018}.

Some existing works have addressed prediction uncertainty only in the context of external disturbances, such as weather predictions \cite{chen2020data, kuijpersWeatherForecastError2022}.
\begin{color}{black}In \cite{garciamanas2024multiscenario}, the uncertainty in market prices is addressed with a scenario-based stochastic MPC controller; however, the controller is based on mixed-integer optimization, requiring significant computational efforts.\end{color}
Alternatively, the following existing works have explored uncertainty stemming from an incorrect physical model of the system.
In \cite{hamzaRobustTSFuzzy2019}, the robustness of predictive control for a greenhouse in the presence of model uncertainties is considered empirically, however no mechanism to compensate for the prediction error is introduced.
\begin{color}{black}In \cite{mahmood2023datadriven, mahmood2023greenhouse}, a neural network is used as a prediction model, with a robust MPC controller addressing the prediction uncertainty; however, a min-max robust MPC approach is used, which is inherently conservative.\end{color}
In \cite{xuAdaptiveTwoTimeScale2018}, an approach to mitigating the negative effects of model uncertainty is presented using online parameter estimation.
However, only a small subset of the (possibly) uncertain model parameters are estimated.
Of particular note, in \cite{boersmaRobustSampleBasedModel2022}, parametric uncertainty in all model parameters is addressed. 
A robust sample-based MPC controller is proposed to incorporate the uncertainty into the control approach. However, the resulting control scheme is unable to adequately reduce constraint violations, and results in less crop yield due to conservativeness. 
More recently, \cite{svensen2024chanceconstrained} has proposed a chance-constrained stochastic MPC formulation to address parametric uncertainties in greenhouse production systems. However, this approach relies on linearization of the prediction model, further increasing prediction uncertainty. Additionally, the chance constrained formulation leads to a computational load that is higher than traditional MPC schemes.

In contrast to MPC, reinforcement learning (RL) is a model-free control methodology where a control policy is learned from data observed from the system \cite{suttonReinforcementLearningIntroduction2018}.
RL controllers naturally handle uncertainty and adapt to changing conditions with no additional mechanisms, as they are learned through direct interaction with the real system.
For complex systems with large continuous state and action spaces, the state-of-the-art for RL uses deep neural networks (DNNs) as function approximators to represent the controller.
With the power of DNNs as general function approximators, this idea has seen unprecedented success on previously unsolved problems, e.g., the games of chess and Go \cite{silverGeneralReinforcementLearning2018}.
The power and inherent uncertainty handling of RL has been identified as useful for greenhouse climate control, {\color{black}with \cite{wangDeepReinforcementLearning2020} proposing a DNN-based RL approach based on deep deterministic policy gradient (DDPG) \cite{lillicrap2015continuous}, and \cite{morcegoReinforcementLearningModel2023b} drawing a comparison between MPC and DDPG}.
However, an inherent drawback for RL controllers is the absence of theoretical guarantees on the satisfaction of constraints and a lack of interpretability due to the black-box DNN function approximation. 
In the context of greenhouse control, this means growers have no guarantee that the automated controller will effectively constrain climate variables to safe ranges, with potential negative implications on crop health and profit.

Recently, \cite{grosDataDrivenEconomicNMPC2020} proposed and justified an integrated MPC and RL control paradigm, where the MPC controller serves as a function approximator for the optimal policy in model-based RL. Figure \ref{introduction:fig:mpcrl-diagram} depicts a schematic overview of this architecture.
In such a scheme, the MPC controller's optimization problem acts both as policy provider, picking actions based on a state, and value function approximator, estimating `how good' it is to be in a given state. 
The learning algorithm, e.g., Q-learning \cite{watkins1992q}, is tasked with adjusting the parametrization of the MPC controller in an effort to discover the optimal control policy, thus improving closed-loop performance in a data-driven fashion. 
In this way, despite the presence of mismatches between the prediction model and the real system, the MPC control scheme is able to learn and deliver, at convergence, the optimal policy and value functions of the underlying RL task, granted the MPC parametrization is rich enough.
In contrast to DNN-based RL, the MPC scheme at the core of this approach provides the option of integrating prior information that may be known on the system in the form of, e.g., an expert-based, possibly imperfect, prediction model. 
Moreover, MPC-based agents are in general more amenable to analysis and certification in terms of stability and constraint satisfaction \cite{borrelliPredictiveControlLinear2016}. 
Finally, MPC-based controllers can take constraints into account in an explicit and structured way, which DNNs are generally incapable of doing.
The above benefits render this methodology suitable for greenhouse climate control, where an inaccurate prediction model is known, and climate variables must be constrained.
\begin{figure}
    \centering
    \begin{tikzpicture}[x=0.75pt,y=0.75pt,yscale=-0.8175,xscale=1]
    \draw (140,89) -- (140,79) -- (212.93,2.18);
    \draw [shift={(215,0)}, rotate = 133.51, fill={rgb,255:red,0;green,0;blue,0}, line width=0.08, draw opacity=0] (10.72,-5.15) -- (0,0) -- (10.72,5.15) -- (7.12,0) -- cycle;
    \draw (109,160) -- (70,160) -- (70,110) -- (91,110);
    \draw [shift={(94,110)}, rotate = 180, fill={rgb,255:red,0;green,0;blue,0}, line width=0.08, draw opacity=0] (10.72,-5.15) -- (0,0) -- (10.72,5.15) -- (7.12,0) -- cycle;
    \draw (109,170) -- (51,170) -- (51,40) -- (96,40) ;
    \draw [shift={(99,40)}, rotate=180, fill={rgb,255:red,0;green,0;blue,0}, line width=0.08, draw opacity=0] (10.72,-5.15) -- (0,0) -- (10.72,5.15) -- (7.12,0) -- cycle;
    \draw (240,40) -- (290,40) -- (290,165) -- (232,165) ;
    \draw [shift={(229,165)}, rotate=360, fill={rgb,255:red,0; green,0;blue,0}, line width=0.08, draw opacity=0] (10.72,-5.15) -- (0,0) -- (10.72,5.15) -- (7.12,0) -- cycle;

    \draw  [color={rgb,255:red,254;green,151;blue,43}, draw opacity=1, fill={rgb,255:red,255;green,230;blue,204}, fill opacity=1] (95,98) .. controls (95,93.58) and (98.58,90) .. (103,90) -- (177,90) .. controls (181.42,90) and (185,93.58) .. (185,98) -- (185,122) .. controls (185,126.42) and (181.42,130) .. (177,130) -- (103,130) .. controls (98.58,130) and (95,126.42) .. (95,122) -- cycle ;
    \draw [color={rgb,255:red,93;green,195;blue,79}, draw opacity=1, fill={rgb,255:red,213;green,232;blue,212}, fill opacity=1] (109,153) .. controls (109,148.58) and (112.58,145) .. (117,145) -- (221,145) .. controls (225.42,145) and (229,148.58) .. (229,153) -- (229,177) .. controls (229,181.42) and (225.42,185) .. (221,185) -- (117,185) .. controls (112.58,185) and (109,181.42) .. (109,177) -- cycle;
    \draw [color={rgb,255:red,87;green,113;blue,214}, draw opacity=1, fill={rgb,255:red,218;green,232;blue,252}, fill opacity=1] (100,28) .. controls (100,23.58) and (103.58,20) .. (108,20) -- (232,20) .. controls (236.42,20) and (240,23.58) .. (240,28) -- (240,52) .. controls (240,56.42) and (236.42,60) .. (232,60) -- (108,60) .. controls (103.58,60) and (100,56.42) .. (100,52) -- cycle;

    \draw (140,110) node [align=left] {\textbf{RL Agent}};
    \draw (169,165) node [align=left] {\textbf{Environment}};
    \draw (170,40) node [align=left] {\textbf{Parametric MPC}};
    \draw (54,82) node [anchor=north west, inner sep=0.75pt, align=left] {state};
    \draw (152,67) node [anchor=north west, inner sep=0.75pt, align=left] {parameters};
    \draw (73,132) node [anchor=north west, inner sep=0.75pt, align=left] {cost};
    \draw (252,82) node [anchor=north west, inner sep=0.75pt, align=left] {action};
\end{tikzpicture}
    \caption{Diagram of the MPC-based RL architecture}
    \label{introduction:fig:mpcrl-diagram}
\end{figure}

Therefore, in this work we propose an integrated MPC and RL framework to address the problem of greenhouse climate control under parametric uncertainty stemming from uncertain weather predictions and modeling mismatches. 
Specifically:
\begin{enumerate}
	\item To the best of the authors' knowledge, a combined MPC and RL approach for greenhouse climate control in the presence of uncertainty is proposed for the first time.
	A parametrized MPC scheme, inspired by \cite{grosDataDrivenEconomicNMPC2020}, is crafted to serve as policy provider and value function approximator in an RL formulation of the greenhouse climate control problem.
	A second-order Q-learning algorithm is leveraged to adjust the parametrization of the MPC scheme online, automatically learning a control policy.
	The approach provides an {\color{black}adaptive} and high-performing climate controller that minimizes potentially dangerous constraint violations, without negatively {\color{black}affecting} crop growth due to robustness conservatism, and without relying on an expensive-to-acquire accurate prediction model. Additionally, in contrast to DNN-based learning approaches, the behavior of the resulting controller can be interpreted by analyzing the learned parametrization of the constraints, prediction model, and cost function. 
	
	\item The approach is then validated in simulation, and compared against both model-based MPC and model-free RL state-of-the-art controllers.
	The results demonstrate the effectiveness of the approach, with the proposed methodology outperforming existing controllers from the literature in both constraint satisfaction and efficient crop growth.
\end{enumerate}

{\color{black} Compared to the state-of-the-art MPC formulations \cite{boersmaRobustSampleBasedModel2022,garciamanas2024multiscenario,mahmood2023datadriven,mahmood2023greenhouse,svensen2024chanceconstrained}, instead of addressing uncertainty in the parametrization in a robust or stochastic fashion, the proposed methodology adapts its policy via RL in order to improve closed-loop performance. This has the distinct advantage of yielding less conservative control schemes while retaining low computational complexity. In comparison with DNN-based approaches such as \cite{morcegoReinforcementLearningModel2023b,wangDeepReinforcementLearning2020}, our method integrates MPC as function approximator in the RL algorithm, fostering a model-based approach that is more suitable for learning high-performance, constraint-abiding policies.}
    
The paper is structured as follows.
In Section \ref{sec:backgound} relevant background is provided on the greenhouse model used, and on the combined MPC and RL paradigm from \cite{grosDataDrivenEconomicNMPC2020}.
The problem we address is formally defined in Section \ref{sec:problem_formulation}.
Section \ref{sec:methodology} presents the proposed methodology, which is then applied and assessed extensively in simulation in Section \ref{sec:simulations}.
Finally, conclusions and future work directions are given in Section \ref{sec:conclusions}.

\section{Background}\label{sec:backgound}
This section describes the greenhouse model considered in this work.
Additionally, background theory on combining MPC and RL is provided, upon which the methodology proposed in this paper is built.

\subsection{Lettuce Greenhouse Model}\label{sec:lettuce_greenhouse_model}
We consider a greenhouse for lettuce growing, with the continuous-time model, presented in \cite{vanhentenGreenhouseClimateManagement1994}, given in the Appendix.
While the continuous-time model is used in all simulations, for control purposes a discrete-time model is considered
\begin{equation}\label{eq:true_model}
	\begin{aligned}
		x(k+1) &= f\big(x(k), u(k), d(k), p\big), \\
		y(k) &= g\big(x(k), p\big), 
	\end{aligned}
\end{equation}
with $x \in \mathbb{R}^4$ the state, $u \in \mathbb{R}^3$ the control input, $d \in \mathbb{R}^4$ the weather disturbance, and $y \in \mathbb{R}^4$ the output.
Furthermore, $p \in \mathbb{R}^{28}$ is a set of model parameters, and $k \in \mathbb{Z}^+$ is the discrete-time counter for discrete time steps of $\Delta t = 900$s (15 minutes).
The nonlinear functions $f$ and $g$, and the model parameters $p$, are given in the Appendix.
The physical meaning of the states, outputs, inputs, and disturbances is summarized in Table~\ref{tab:state}.
Estimation is out the scope of this paper, and it is assumed, as in \cite{boersmaRobustSampleBasedModel2022}, that at each time step $k$ a perfect estimate of the state $x(k)$ is available.
\begin{table*}[ht]
    \caption{Physical meaning of the state $x$, output $y$, input $u$, and disturbance $d$.}
    \begin{center}
    	\begin{tabular}{clclcl}
    		\toprule 			
    		$x_{1}$ & dry-weight (\unit{\kilogram\per\square\meter})                     & 
            $y_{1}$ & dry-weight (\unit{\gram\per\square\meter})                         &
            $d_{1}$ & radiation (\unit{\watt\per\square\meter})                          \\
    		$x_{2}$ & indoor CO$_2$ (\unit{\kilogram\per\cubic\meter})                   & 
            $y_{2}$ & indoor CO$_2$ (\unit{\permille})                                   & 
            $d_{2}$ & outdoor CO$_2$ (\unit{\kilogram\per\cubic\meter})                  \\
    		$x_{3}$ & indoor temperature (\unit{\celsius})                               & 
            $y_{3}$ & indoor temperature (\unit{\celsius})                               & 
            $d_{3}$ & outdoor temperature (\unit{\celsius})                              \\
    		$x_{4}$ & indoor humidity (\unit{\kilogram\per\cubic\meter})                 &
            $y_{4}$ & indoor humidity (\unit{\percent})                                  & 
            $d_{4}$ & outdoor humidity (\unit{\kilogram\per\cubic\meter})                \\
    		\midrule	    
    		$u_{1}$ &  CO$_2$ injection (\unit{\milli\gram\per\square\meter\per\second}) &
            $u_{2}$ &  ventilation  (\unit{\milli\meter\per\second})                     &
            $u_{3}$ &  heating (\unit{\watt\per\square\meter})                           \\
    		\bottomrule
    	\end{tabular}
    	\label{tab:state}
    \end{center}
\end{table*}

\subsection{MPC as a function approximator in RL}\label{sec:mpc_func_approx}
Consider discrete-time system dynamics as a Markov Decision Process (MDP) \cite{suttonReinforcementLearningIntroduction2018} with continuous state $s \in \mathbb{R}^{n}$, continuous action $a \in \mathbb{R}^{m}$, and state transitions $s,a \rightarrow s_+$ with the underlying conditional probability density
\begin{equation} \label{eq:background:rl:transition_probab}
	\mathbb{P}[s_+ | s, a] : \mathbb{R}^{n} \times \mathbb{R}^{m} \times \mathbb{R}^{n} \rightarrow [0, 1].
\end{equation}
Consider a deterministic policy $\pi_\theta(s) : \mathbb{R}^{n} \rightarrow \mathbb{R}^{m}$ parametrized by $\theta \in \mathbb{R}^{l}$.
Selecting actions based on this policy will cause the system to visit the MDP's states with a given distribution, denoted $\eta_{\pi_\theta}$. The performance of such a policy is defined as \cite{suttonReinforcementLearningIntroduction2018}
\begin{equation}  \label{eq:RL_performance}
	J(\pi_\theta) = \mathbb{E}_{\eta_{\pi_\theta}} \left[
        \sum_{k=0}^{N_\text{s}}{\gamma^k L \Bigl(s_k, \pi_\theta\big(s_k\big)\Bigr)}
    \right],
\end{equation}
where $s_k$ is the state at time step $k$, $L(s,a) : \mathbb{R}^{n} \times \mathbb{R}^{m} \rightarrow \mathbb{R}$ the stage cost, $\gamma \in (0, 1]$ the discount factor, and $N_\text{s}$ the number of time steps considered in a task. The RL task is then to find the optimal policy $\pi^\star_\theta$ as
\begin{argmini}
    { \theta }{ J(\pi_\theta) }{ \label{eq:RL_optimal_pol} }{ \pi^\star_\theta = }.
\end{argmini}
The familiar notions of state- and action-value functions \cite{suttonReinforcementLearningIntroduction2018} are defined respectively as
\begin{multline}
	\label{eq:val_func_defs}
     Q_\theta\big(s_k, a_k\big) = L\Big(s_k, a_k\Big) \\ 
     + \mathbb{E}_{\eta_{\pi_\theta}}\left[\sum_{\tau = k+1}^{N_\text{s}} \gamma^{\tau - k} L \Bigl(s_k, \pi_\theta\big(s_k\big)\Bigr)\right],
\end{multline}
and $V_\theta\big(s_k\big) = Q_\theta\Big(s_k, \pi_\theta\big(s_k\big)\Big)$.
While DNNs are the most common choice for representing the policy and value functions \cite{lillicrap2015continuous}, their black-box nature does not facilitate the injection of prior information, e.g., approximate prediction models, nor is it conducive to an interpretabile policy and the addition of constraints.   
To account for these drawbacks, \cite{grosDataDrivenEconomicNMPC2020} proposed the use of an MPC scheme in place of a DNN.

Consider the following MPC problem approximating the value function, parametrized by $\theta$, $V_\theta : \mathbb{R}^{n} \rightarrow \mathbb{R}$ as
\begin{mini!}
    { \textbf{x}, \textbf{u}, \bm{\sigma} }{
        \lambda_\theta\big(x(0)\big) 
        \nonumber 
        \label{eq:gen_MPC_cost}
    }{ \label{eq:gen_MPC} }{ V_\theta(s) = }
    \breakObjective{
        \hspace{3.3em} 
        + \sum_{k=0}^{N-1} \gamma^k \Big(L_\theta\big(x(k), u(k)\big) + \omega^\top \sigma(k)\Big) 
        \nonumber
    }
    \breakObjective{
        \hspace{2.5em} 
        + \gamma^N\Big(V_\theta^\text{f}\big(x(N)\big) + w_\text{f}^\top\sigma(N)\Big)
    }
    \addConstraint{ x(0) = s, \label{eq:first_constraint} }
    \addConstraint{ \text{for} \enspace k = 0,\dots,N-1 \nonumber }
    \addConstraint{ \enspace x(k+1) = f_\theta\big(x(k), u(k)\big), }
    \addConstraint{ \enspace h_\theta\big(x(k), u(k)\big) \leq \sigma(k), }
    \addConstraint{ \enspace \sigma(k) \geq 0, }
    \addConstraint{ h_\theta^\text{f}\big(x(N)\big) \leq \sigma(N), }
    \addConstraint{ \sigma(N) \geq 0, \label{eq:last_constraint} }
\end{mini!}
where slack variable $\sigma(k)$ softens the inequality constraint for time step $k$, and the vectors $\textbf{x}$, $\textbf{u}$ and $\bm{\sigma}$ respectively collect the states, actions, and slack variables over the horizon $N$. 
Problem \eqref{eq:gen_MPC} is solved numerically online to generate the value $V_\theta(s)$.
In \eqref{eq:gen_MPC}, $\lambda_\theta$ is an initial cost term, $L_\theta$ is the stage cost, and $V_\theta^\text{f}$ is a terminal cost approximation, all of which are parametrized by $\theta$.
Furthermore, $f_\theta$ is the model approximation, and $h_\theta$, $h^\text{f}_\theta$ are inequality constraints. Lastly, $w$ and $w_{\text{f}}$ are the weights of the slack variable in the objective. 
\begin{color}{black}Note that the formulation \eqref{eq:gen_MPC} is general, i.e., the dimension of $\theta$, and how it enters into the respective functions in \eqref{eq:gen_MPC}, is not made explicit. In Section~\ref{sec:methodology} we will introduce a concrete realization, designed for the greenhouse climate control problem.\end{color} 

Given \eqref{eq:gen_MPC}, the action value function $Q_\theta$ and the policy $\pi_\theta$, satisfying the fundamental equalities of the Bellman equations \cite{grosDataDrivenEconomicNMPC2020}, are defined as follows:
\begin{mini}
    { \textbf{x}, \textbf{u}, \bm{\sigma} }{ 
        \eqref{eq:gen_MPC_cost} 
    }{}{ Q_\theta(s,a) = }
    \addConstraint{ \eqref{eq:first_constraint}-\eqref{eq:last_constraint}, }
    \addConstraint{ u(0) = a, }
\end{mini}
\begin{argmini}
    { a }{ Q_\theta(s,a) }{}{ \pi_\theta(s) = }.
\end{argmini}
Therefore, in RL terms, the parametric MPC scheme acts as policy provider for the learning agent, whose goal is to modify the parameters $\theta$ of the controller in order to minimize \eqref{eq:RL_optimal_pol}. Various forms of RL \cite{suttonReinforcementLearningIntroduction2018} exist that solve this problem directly or indirectly via iterative updates
\begin{equation} \label{eq:background:rl:update}
	\theta \leftarrow \theta - \alpha \nabla_\theta \sum_{i=1}^{m}{\psi(s_i,a_i,s_{i+1},\theta)},
\end{equation}
where $\alpha \in \mathbb{R}_+$ is the learning rate, $z$ denotes the number of observations used in the update (i.e., a batch of observations), and $\psi$ captures the controller's performance and varies with the specific algorithm. For example, in recursive Q-learning we have that $m=1$ and $\psi = \delta_i^2$, where the temporal-difference (TD) error 
\begin{equation}
	\delta_i = L(s_i,a_i) + \gamma V_\theta(s_{i+1}) - Q_\theta(s_i,a_i)
\end{equation}
captures the estimation error of the value functions \cite{suttonReinforcementLearningIntroduction2018}.

\section{Problem Formulation}\label{sec:problem_formulation}
We address the problem of optimal greenhouse climate control for crop yield and resource efficiency.
As in \cite{boersmaRobustSampleBasedModel2022}, we consider the predictive uncertainty, stemming from uncertain weather predictions and modeling errors, to be captured by parametric uncertainty\footnote{\begin{color}{black}The choice to represent modeling error and weather prediction inaccuracy via the uncertainty in $p$ results in a prediction model in which even the parameters that may represent known physical quantities, e.g., the ideal gas constant $p_{23}$, being unknown.\end{color}} in the model parameters $p$. 
Specifically, the true values for $p$ are assumed to be unknown.

\begin{color}{black}Growth cycles of 40 days are considered, where control inputs are computed at 15 minute time steps, i.e., growth cycles of $N_\text{s} = 3840$ time steps.
This duration is in line with the literature \cite{boersmaRobustSampleBasedModel2022, morcegoReinforcementLearningModel2023b}, and represents a standard lettuce growth cycle, at the end of which the lettuce is harvested and sold, generating economic profit.\end{color}
During each growth cycle, we wish to maximize the yield and minimize the violations of constraints on the system outputs, whilst minimizing the {\color{black}cost associated with} control signals.
The system outputs are constrained as $y_\text{min}(k) \leq y(k) \leq y_\text{max}(k)$, with \cite{boersmaRobustSampleBasedModel2022, morcegoReinforcementLearningModel2023b}
\begin{equation}
	\begin{aligned}
		y_\text{min}(k) &= \Big(0, 0, y_{3, \text{min}}\big(d_1(k)\big), 0\Big)^\top, \\ 
		y_\text{max}(k) &= \Big(\infty, 1.6, y_{3, \text{max}}\big(d_1(k)\big), 70\Big)^\top,
	\end{aligned}
\end{equation}
and with the time-varying third element defined as
\begin{equation}
	\begin{aligned}
		y_{3, \text{min}}\big(d\big) &= \begin{cases}
			10, & \text{if } d < 10 \\
			15, & \text{if } d \geq 10
		\end{cases}, \\
		y_{3, \text{max}}\big(d\big) &= \begin{cases}
			15, & \text{if } d < 10 \\
			20, & \text{if } d \geq 10
		\end{cases}.
	\end{aligned}
\end{equation}
These time varying constraints are common in the literature; they reflect that inside the greenhouse it is colder during the night than during the day \cite{morcegoReinforcementLearningModel2023b, boersmaRobustSampleBasedModel2022, seginer1994optimal}.
The inputs are constrained as $u_\text{min} \leq u(k) \leq u_\text{max}$, with \cite{boersmaRobustSampleBasedModel2022, morcegoReinforcementLearningModel2023b}
\begin{equation}
	\begin{aligned}
		u_\text{min} &= (0, 0, 0)^\top, \\
		u_\text{max} &= (1.2, 7.5, 150)^\top. \\
	\end{aligned}
\end{equation}
Finally, the input rate is constrained as \cite{boersmaRobustSampleBasedModel2022, morcegoReinforcementLearningModel2023b}
\begin{equation} \label{eq:control_rate_cnstr}
    |u(k+1)-u(k)| \leq (1/10) u_\text{max}.
\end{equation}

Define the following performance indicators:
\begin{itemize}
	\item Final yield $y_1(N_\text{s})$
	\item Constraint violations $\Psi$
	\item Economic profit indicator $P$.
\end{itemize}
The final yield $y_1(N_\text{s})$ is the dry lettuce weight at the end of a growth cycle of $N_\text{s}$ time steps.
The constraint violations indicator $\Psi$ is defined as
\begin{multline}
    \Psi = \sum_{k=0}^{N_\text{s}} \sum_{i=1}^4 \Bigg(\max\bigg(0, \frac{y_i(k) - y_{i, \text{max}}(k)}{y_{i, \text{max}}(k) - y_{i, \text{min}(k)}}\bigg) \\
    + \max\bigg(0, \frac{y_{i, \text{min}}(k) - y_i(k)}{y_{i, \text{max}}(k) - y_{i, \text{min}}(k)}\bigg)\Bigg).
\end{multline}
This performance indicator captures the magnitude of violations on the output constraints over a growth cycle.
Finally, the economic profit indicator\footnote{\begin{color}{black}Note that $P$ represents a profit indicator per square meter, as the lettuce model in this work is normalized by the surface area (see Section~\ref{sec:lettuce_greenhouse_model} and \cite{vanhentenGreenhouseClimateManagement1994,vanhenten2003sensitivity}).\end{color}} $P$ is defined as \cite{vanhenten2003sensitivity,morcegoReinforcementLearningModel2023b}
\begin{multline}
    P = c_{\text{price}, 1} + c_{\text{price}, 2} \: y_1(N_\text{s}) \\
    - \sum_{k=0}^{N_\text{s}} \big(c_{\text{CO}_2}u_1(k) + c_\text{q}u_3(k)\big) \Delta t,
\end{multline}
{\color{black}where the relation between auction price and harvest weight of lettuce is modeled linearly with coefficients $c_{\text{price}, 1}$ and $c_{\text{price}, 2}$, and the financial cost of the climate conditioning equipment is linearly related to the amount of energy and carbon dioxide put into the system, weighted by prices $c_\text{q}$ and $c_{\text{CO}_2}$, respectively. Note that in $P$ no cost is associated to natural ventilation used for cooling and dehumidification, i.e., $u_2$ \cite{garciamanas2024multiscenario}. The values of the coefficients of this economic model are given in Table~\ref{tab:EPI}, and more details are available in, e.g., \cite[Section~2.1]{vanhenten2003sensitivity}.}
This performance indicator represents the monetary value of a growth cycle, and captures the efficiency of the control with respect to the use of costly actuators.
\begin{table}
	\caption{Coefficient values for the economic profit indicator $P$ (\begin{color}{black}The former Dutch currency Hfl, the Guilder, is used for adherence to the literature \cite{boersmaRobustSampleBasedModel2022, morcegoReinforcementLearningModel2023b, svensen2024chanceconstrained, vanhentenGreenhouseClimateManagement1994, vanhenten2003sensitivity}\end{color}).}
	\begin{center}
		\begin{tabular}{lcl}
			\toprule
            Symbol           & Value                & Unit  \\
            \midrule
			$c_{\text{price}, 1}$ & $1.8$                & \unit{\hollandseflorijn\per\square\meter}  \\
			$c_{\text{price}, 2}$ & $1.6$                & \unit{\hollandseflorijn\per\kilogram}  \\
            $c_\text{q}$          & $6.35 \cdot 10^{-9}$ & \unit{\hollandseflorijn\per\joule}  \\
            $c_{\text{CO}_2}$     & $42 \cdot 10^{-2}$   & \unit{\hollandseflorijn\per\kilogram}  \\
			$\Delta t$            & $900$                & \unit{\second}  \\
			\bottomrule
		\end{tabular}
		\label{tab:EPI}
	\end{center}
\end{table}
\section{Methodology}\label{sec:methodology}
To address the issues caused by inaccurate knowledge of the true prediction model parameters $p$ for model-based MPC controllers, we propose a novel parametrized MPC scheme for the greenhouse climate control problem.
Then, leveraging closed loop data, we show how the parameter values can be learned using RL techniques \cite{grosDataDrivenEconomicNMPC2020}, compensating for the performance loss introduced by an inaccurate prediction model.

\subsection{Greenhouse Climate Control as an RL Problem}\label{seq:green_RL_prob}
In order to apply an RL methodology to learn the MPC parametrization, the greenhouse climate control must be modeled as an RL task, i.e., as an MDP defined as in Section~\ref{sec:mpc_func_approx}. 
Trivially, the greenhouse input variable $u$ corresponds directly to the actions $a$ in the RL context.
Define the concatenation of the current state, and the current and previous outputs, as the RL state:
\begin{equation}
	s(k) = \big(x^\top(k), y^\top(k), y^\top(k-1)\big)^\top.
\end{equation}
Furthermore, the state transitions are determined by the true system model \eqref{eq:true_model}, where the exact probabilities are conditioned on the weather disturbance $d$.
Note that, to enforce the control rate constraint \eqref{eq:control_rate_cnstr} in the MDP, the input variable is clipped prior to being applied to the system.

The stage cost $L$ requires particular attention, as it implicitly defines the optimization problem \eqref{eq:RL_optimal_pol} that the RL agent is tasked with solving.
Crafting the stage cost $L$ such that the resulting RL policy performs well with respect to the performance indicators defined in Section \ref{sec:problem_formulation} is a design challenge.
Consider the stage cost
\begin{equation}
	L\big(s(k), a(k)\big) = L_{y_1}\big(s(k)\big) + L_{u}\big(a(k)\big) + L_{\Psi}\big(s(k)\big).
\end{equation}
The function
\begin{equation}
	L_{y_1}\big(s(k)\big) = -c_{\delta y_1} \big(y_1(k) - y_1(k-1)\big),
\end{equation}
with $c_{\delta y_1} > 0$, rewards the step-wise increase in dry lettuce weight.
The function 
\begin{equation}
	L_{u}\big(a(k)\big) = c_u^\top u(k),
\end{equation} 
with $c_u \in \mathbb{R}^3 > 0$, penalizes the control inputs, while the function
\begin{multline}
    L_{\Psi}\big(s(k)\big) = \omega^\top \max \left(
        0, \frac{y(k) - y_\text{max}(k)}{y_\text{max}(k) - y_\text{min}(k)}
    \right) \\
    + \omega^\top \max \left(
        0, \frac{y_\text{min}(k) - y(k)}{y_\text{max}(k) - y_\text{min}(k)}
    \right)
\end{multline}
penalizes constraint violations, with $\omega \in \mathbb{R}^4 > 0$, and with the max operator and vector division performed element-wise.
The constants $c_{\delta y_1}$, $c_u$, and $\omega$ are then hyper-parameters that are tuned such that the RL policy performs well on the performance indicators in Section~\ref{sec:problem_formulation}.
Lastly, the RL policy $\pi_\theta$, and value functions $V_\theta$ and $Q_\theta$, are addressed via the parametrized MPC scheme, which is introduced next.

\subsection{Parametrized MPC Scheme}
In this section we introduce the parametrized MPC scheme that, following the theory outlined in Section~\ref{sec:mpc_func_approx}, serves as policy provider and value function approximator for the RL task outlined in Section~\ref{seq:green_RL_prob}.

\begin{color}{black}Consider the following parametrized MPC scheme, a concrete realization of \eqref{eq:gen_MPC}, representing the value function, with prediction horizon $N \in \mathbb{Z}^+$ and discount factor $\gamma \in (0, 1]$\end{color}:
\begin{subequations}
	\label{eq:MPC_parametrized}
	\begin{align}
		V_\theta(s) = 
        & \min_{\displaystyle \textbf{y}, \textbf{x}, \textbf{u}, \bm{\sigma}} 
        \enspace \theta_0 - \theta_{\delta y_1}\sum_{k=1}^N \gamma^k \big(y_1(k) - y_1(k-1) \big) \nonumber \\
        & \hspace{4.2em}
        + \theta_u^\top\sum_{k=0}^{N-1} \gamma^k u(k)
        + \theta_\omega^\top \sum_{k=0}^N \gamma^k \frac{\sigma(k)}{y_\text{range}} \nonumber\\
        & \hspace{4.2em}
        + \theta_{y_1, \text{f}} \: \gamma^N \: V_{\theta, \text{f}}\big(y_1(N)\big) \label{eq:cost} \\
        \text{s.t.} \qquad 
        & x(0) = x, \label{eq:IC} \\
        & \text{for} \enspace k = 0,\dots,N-1 \nonumber \\
        & \enspace x(k+1) = f\big(x(k), u(k), d(k), \theta_p\big), \label{eq:dynamics} \\
        & \enspace u_\text{min} \leq u(k) \leq u_\text{max}, \label{eq:control_cnstrs} \\
        & \enspace -\delta u \leq u(k) - u(k-1) \leq \delta u, \label{eq:control_rate_cnstrs} \\
        & \text{for} \enspace k = 0,\dots,N \nonumber \\
        & \enspace y(k) = g\big(x(k), \theta_p\big), \label{eq:outputs} \\
        & \enspace y_\text{min}(k) - \sigma(k) \leq y(k) \leq y_\text{max}(k) + \sigma(k), \label{eq:output_cnstrs} \\
        & \enspace \sigma(k) \geq 0, \label{eq:slack_cnstrs}
	\end{align}
\end{subequations}
where\footnote{The $\infty$ in $y_\text{range}$ causes the first component of the violation penalty to be always zero, as the first output is unbounded.} $y_\text{range} = (\infty, 1.6, 5, 70)^\top$.
The cost \eqref{eq:cost} rewards step-wise lettuce weight increase, and penalizes constraint violations and control inputs.
The first and last terms, $\theta_0$ and $V_{\theta, \text{f}}$, are additional {\color{black}initial} offset and terminal costs respectively, enriching the parametrization to help the MPC scheme capture the true RL value functions. {\color{black}In particular, $\theta_0$ is required in order to learn the (possibly non-zero) offset in the value function due to its economic nature (see \cite{grosDataDrivenEconomicNMPC2020} for more theoretical details), while $V_{\theta, \text{f}}$ encodes the cost-to-go.}
The parameters $\theta_{\delta_{y_1}}, \theta_u, \theta_\omega$, and $\theta_{y_{1, \text{f}}}$ weight the contributions of each cost term. 
Since in general $\theta_p \neq p$, the constraints for the dynamics and output equations, \eqref{eq:dynamics} and \eqref{eq:outputs}, are parametrized by $\theta_p$ in place of the true model parameters $p$. 
Hence, the predicted state and output trajectories from solving \eqref{eq:MPC_parametrized} do not necessarily respect the dynamics and output functions of the true system.
The parameter $\theta_p$ then provides a degree of freedom to optimize the predicted trajectories for closed-loop performance, using closed-loop data.
\begin{remark}
	Note that the MPC parameters $\theta_{\delta y_1}$, $\theta_u$, $\theta_\omega$, and $\theta_p$ are distinct from the parameters in the RL stage cost $c_{\delta y_1}$, $c_u$, $\omega$, and the true model parameters $p$.
	While they represent equivalent notions, the MPC parameters will be adjusted to optimize closed-loop performance and will not, in general, converge to the corresponding values in the RL stage-cost and the true model.
    \begin{color}{black}Indeed, in \cite{grosDataDrivenEconomicNMPC2020} it is stressed that the prediction model yielding the best closed-loop performance for an MPC controller is not necessarily the one with the smallest prediction error.
    Hence, even if instantiated with accurate model parameters $\theta_p = p$, it is likely the RL agent would adjust $\theta_p$ regardless, optimizing for closed-loop performance.\end{color}
\end{remark}

The terminal cost $V_{\theta, \text{f}}$ is added to capture the future reward from lettuce weight increase for the remainder of the growth cycle, occurring after the prediction horizon of the MPC controller.
The step-wise reward for lettuce growth for the remainder of the cycle can be simplified as \begin{color}{black}
\begin{align}  
    \begin{split}
        \sum_{k=N+1}^{N_\text{s}} \gamma^k & L_{y_1}\big(s(k)\big)  \\
        ={}& -c_{\delta y_1} \sum_{k=N+1}^{N_\text{s}}  \gamma^k \big(y_1(k) - y_1(k-1) \big) 
    \end{split}  \\
    \begin{split}
        ={}& -c_{\delta y_1} \Biggl( \gamma^{N_\text{s}} y_1(N_\text{s}) - \gamma^{N+1} y_1(N)  \\
        {}&+ \sum_{k=N+1}^{N_\text{s}-1} \left( \gamma^k -  \gamma^{k+1}\right) y_1(k) \Biggr),
    \end{split}  \\
    \approx{}& -c_{\delta y_1} \bigl( y_1(N_\text{s}) - y_1(N) \bigl),\label{eq:tel_sum}
\end{align}
where, for the sake of simplicity, it is assumed that $\gamma \approx 1$, which is common when a non-myopic policy is desired \cite{suttonReinforcementLearningIntroduction2018}. Of course, \eqref{eq:tel_sum} cannot be used as terminal cost directly, as the final yield $y_1(N_\text{s})$ is not known and varies depending on the weather disturbance and the control inputs.
As the value functions attempt to capture the expected cost under the current policy (see \eqref{eq:val_func_defs}), we introduce the learnable parameter $\theta_{y_N}$ in order to learn the expected value $\mathbb{E}_{\eta_{\pi_\theta}}\left[y_1(N_\text{s})\right]$, where, again, $\eta_{\pi_\theta}$ is the state distribution induced by the policy $\pi_\theta$ and the weather disturbance $d$.
To capture \eqref{eq:tel_sum},\end{color} the terminal cost is then defined as
\begin{equation}
	V_{\theta, \text{f}}\big(y_1(N)\big) = -\theta_{\delta y_1}\big(\theta_{y_N} - y_1(N)\big).
\end{equation}
The full parametrization is then
\begin{equation}
	\theta = (\theta_0, \theta_{\delta y_1}, \theta_u^\top, \theta_\omega^\top, \theta_{y_1, \text{f}}, \theta_{y_N}, \theta_p^\top)^\top.
\end{equation}
The allowable range for the parameters $\theta$ can be bounded during learning, incorporating prior knowledge on viable ranges, or enforcing realistic values on parameters that have physical significance. 
Table~\ref{tab:parameterization} gives a summary of the parametrization, while the bounds and initialization of the parameters in our experiments is given in Section \ref{sec:simulations}. 
\begin{table}
	\caption{Summary of MPC parametrization in \eqref{eq:MPC_parametrized}.}
	\begin{center}
		\begin{tabular}{llc}
			\toprule 	
			Symbol                   & Scope                              & Space  \\
			\midrule		
			$\theta_0$               & cost - offset                      & $\mathbb{R}$  \\
			$\theta_{\delta y_1}$    & cost - weight reward               & $\mathbb{R}$  \\
			$\theta_u$               & cost - control penalty             & $\mathbb{R}^3$  \\
			$\theta_\omega$          & cost - violations penalty          & $\mathbb{R}^4$  \\
			$\theta_{y_1, \text{f}}$ & cost - terminal cost weight        & $\mathbb{R}$  \\
			$\theta_{y_N}$           & cost - terminal weight estimate    & $\mathbb{R}$  \\
			$\theta_p$               & model parameters                   & $\mathbb{R}^{28}$  \\
			\bottomrule
		\end{tabular}
		\label{tab:parameterization}
	\end{center}
\end {table}

As outlined in Section~\ref{sec:mpc_func_approx}, the MPC scheme \eqref{eq:MPC_parametrized} for the value function approximation $V_\theta(s)$ satisfies the Bellman equalities \cite{grosDataDrivenEconomicNMPC2020}, such that the action-value function and policy are delivered by the same scheme as
\begin{mini}
    { \textbf{y}, \textbf{x}, \textbf{u}, \bm{\sigma} }{ 
        \eqref{eq:cost} 
    }{ \label{eq:Q_func} }{ Q_\theta(s,a) = }
    \addConstraint{ \eqref{eq:IC}-\eqref{eq:slack_cnstrs}, }
    \addConstraint{ u(0) = a, }
\end{mini}
\begin{argmini}
    { u }{ Q_\theta(s, u) }{ \label{eq:bellman_policy} }{ \pi_\theta(s) = }.
\end{argmini}
Instead of \eqref{eq:bellman_policy}, the equivalent form
\begin{argmini}
    { u(0) }{ \eqref{eq:cost} }{}{ \pi_\theta(s) = }
    \addConstraint{ \eqref{eq:IC}-\eqref{eq:slack_cnstrs}, }
\end{argmini}
is used in practice such that both $V_\theta(s)$ and $\pi_\theta(s)$ are found by solving the one optimization problem \eqref{eq:MPC_parametrized}.

\subsection{Second-Order LSTD Q-learning} \label{sections:sub:2nd-lstd-q-learning}
We apply a second-order least-squares temporal difference (LSTD) Q-learning algorithm \cite{lagoudakis_2002_leastsquares} to learn the parametrization $\theta$ of \eqref{eq:MPC_parametrized}, such that the policy $\pi_\theta$ optimizes the greenhouse climate control RL task, as defined in Section~\ref{seq:green_RL_prob}. 
Q-learning attempts to learn a parametrization $\theta$ such that the action-value function $Q_\theta$ fits the observed data.
The policy is then inferred from the action-value function as $\pi_\theta(s) = \text{arg}\min_a Q_\theta(s, a)$.
\begin{color}{black}In this work the Q-learning algorithm is applied online, i.e., training data is generated via interaction with the real system, as the policy is trained.\end{color}
Reformulating the fitting of $Q_\theta$ as a least-squares problem, in combination with a second-order Newton's method and an experience replay buffer of the past observed transitions \cite{lin_1992_selfimproving}, provides faster convergence and better sample efficiency with respect to traditional first-order methods.
This approach has been effectively applied in the context of MPC \cite{esfahaniApproximateRobustNMPC2021, airaldi2023reinforcement}.

As in \cite{airaldi2023reinforcement}, we impose lower and upper bounds on the parameters with the following constrained optimization problem
\begin{argmini}
    { \Delta \theta }{ 
        \frac{1}{2}\Delta \theta^\top \bar{H} \Delta \theta + \alpha \bar{g}^\top \Delta \theta
    }{\label{eq:constrained_update}}{ \Delta \theta^\ast = }
    \addConstraint{ \theta_\text{lb} \leq \theta + \Delta \theta \leq \theta_\text{ub}, }
    \addConstraint{ \Delta \theta_\text{lb} \leq \Delta \theta \leq \Delta \theta_\text{ub}, }
\end{argmini}
{\color{black}with $\alpha > 0$ the learning rate, and $\bar{g}$ and $\bar{H}$ the gradient and Hessian of the Q-learning fitting problem averaged over $m$ samples drawn from the replay buffer, i.e.,
\begin{align}
    \bar{g} ={}& -\sum_{i=1}^m  \delta_i \nabla_\theta Q_\theta(s_i, a_i),  \\
    \bar{H} ={}& \sum_{i=1}^m \nabla_\theta Q_\theta(s_i, a_i) \nabla_\theta Q_\theta^\top(s_i, a_i) - \delta_i \nabla_\theta^2 Q_\theta(s_i, a_i).
\end{align}
Note that the sensitivities $\nabla_\theta Q_\theta$ and $\nabla_\theta^2 Q_\theta$ are available automatically upon solving \eqref{eq:Q_func}; see \cite{airaldi2023reinforcement,esfahaniApproximateRobustNMPC2021} for details. } This formulation additionally allows to limit the rate of change of each parameter with $\Delta\theta_\text{lb}$ and $\Delta\theta_\text{ub}$. 
Finally, the parametrization $\theta$ is then updated as $\theta \leftarrow \theta + \Delta\theta^\star$.

\begin{color}{black}Note that during training of the policy two optimization problems are solved at each time step: \eqref{eq:MPC_parametrized} for $V_{\theta}$ and $\pi_\theta$, and \eqref{eq:Q_func} for $Q_\theta$, in order to be able to compute and store in buffer the TD error and sensitivities necessary for an update.
Furthermore, the quadratic program \eqref{eq:constrained_update} must be solved only once per parameter update.
In contrast, when the policy is not being trained, only \eqref{eq:MPC_parametrized} must be solved for $\pi_\theta$.\end{color}

\begin{remark}
    \color{black}
    In general, problem \eqref{eq:RL_optimal_pol} is nonlinear and non-convex. This implies that the RL algorithm is likely to converge to local suboptimal solutions. To alleviate this issue, exploratory behavior can be injected into the learning policy in an effort to escape such local minima and to converge to a better (possibly global) solution. One method is to perturb the gradient of the objective of $Q_\theta$ in an, e.g., $\varepsilon$-greedy fashion \cite{airaldi2023reinforcement}. In the current work, exploration is instead induced during the learning process by the weather forecast disturbances which, as explained in more detail in Section \ref{sec:weather-disturbances}, are stochastically generated at the beginning of each training episode. As later shown in Section \ref{sec:results}, this is enough to provide satisfactory convergence of the TD error."
\end{remark}

\section{Numerical Experiments}\label{sec:simulations}
In this section the methodology proposed in Section~\ref{sec:methodology} is demonstrated in simulation, with the resulting performance compared against existing state-of-art approaches.
In the following, all MPC simulations are run on a Linux machine using one AMD EPYC 7252 core, 1.38GHz clock speed, and 251Gb of RAM.
All simulations for training DNN-based RL methods, i.e., DDPG, are run on the same Linux machine using four NVIDIA RTX 2090 GPUs.
Python source code and simulation results can be found at \url{https://github.com/SamuelMallick/mpcrl-greenhouse}.
All optimization problems are solved using the CasADi framework \cite{andersson2019casadi} and the IPOPT solver \cite{wachter2006implementation}.

\begin{color}{black}Assume that the values of all model parameters $p$ are unknown.
What is known is an uncertainty range $\big[p - 0.2\cdot|p|, \ p + 0.2\cdot|p|\big]$ that contains the true values\end{color}, where $|\cdot|$ is the element-wise absolute value, and defines a uniform distribution spanning the uncertainty range as
\begin{equation}
	\mathcal{R} = \mathcal{U}\big(p - 0.2\cdot|p|, \ p + 0.2\cdot|p|\big).
\end{equation}
The initial values for the learnable model parameters are then a random sample from $\mathcal{R}$, and they are bounded to an enlargement of the uncertainty range.
To initialize the cost parameters, the values from the RL stage cost are used.
Further, we bound all cost terms to be non-negative, such that the notion of rewarded and penalized behavior does not invert.
The initial values and bounds for the full MPC parametrization is given in Table \ref{tab:initialization}.
The RL stage cost coefficients used in our experiments are given in Table~\ref{tab:RL_coeffs}.
\begin{table}
	\caption{Initial values and bounds on parametrization in \eqref{eq:MPC_parametrized}.}
	\begin{center}
		\begin{tabular}{lcc}
			\toprule 	
			Symbol                   & Initial value                   & Bounds  \\
			\hline		
			$\theta_0$               & $0$                             & $(-\infty, \infty)$  \\
			$\theta_{\delta y_1}$    & $100$                           & $(0, \infty)$  \\
			$\theta_u$               & $(10, 1, 1)^\top$               & $(0, \infty)$  \\
			$\theta_\omega$          & $(10^5, 10^5, 10^5, 10^5)^\top$ & $(0, \infty)$  \\
			$\theta_{y_1, \text{f}}$ & $1$                             & $(0, \infty)$ \\
			$\theta_{y_N}$           & $135$                           & $(0, \infty)$\\
			$\theta_p$               & $\sim \mathcal{R}$              & $(0.5|p|, 1.5|p|)$\\
			\bottomrule
		\end{tabular}
		\label{tab:initialization}
	\end{center}
\end{table}
\begin{table}
	\caption{Coefficients in the RL stage cost.}
	\begin{center}
		\begin{tabular}{lc}
			\toprule 			
			Symbol        & Value  \\
			\hline
			$c_{\delta y_1}$ & $100$  \\
			$c_u$            & $(10, 1, 1)^\top$  \\
			$\omega$         & $10^5$\\
			\bottomrule
		\end{tabular}
		\label{tab:RL_coeffs}
	\end{center}
\end{table}

\subsection{Weather Disturbances}  \label{sec:weather-disturbances}

The weather disturbance profile $d$ used in the experiments is real weather data presented in \cite{kempkes2013greenhouse}, collected from `the Venlow Energy greenhouse', located in Bleiswijk, The Netherlands. The data covers a full growth cycle of 40 days and, originally sampled at 5 minute intervals, has been resampled using an FIR anti-aliasing low pass filter at the sampling time used in this work, i.e., 15 minutes \cite{boersmaRobustSampleBasedModel2022}.

As RL approaches train on repeated growth cycle episodes, to avoid over-fitting of the learned policy to a specific instance of weather data and {\color{black}to foster generalization over a wider range of weather disturbances}, a stochastic process is added {\color{black}to the real weather data in order to generate a new perturbed profile for each episodic growth cycle}. In particular, Brownian noise, a time-correlated stochastic process, is added to the original weather profile. The Brownian noise is generated as the cumulative sum of white noise as
\begin{equation}
	B(k) = \sum_{\tau=0}^k W(\tau), \: W(\tau)\sim \mathcal{U}(-\rho, \rho),
\end{equation}
where $\rho=(0.01, 0.005, 0.01, 0.005)^\top$, and $B(k)$ is the noise value at time step $k$.
For the outdoor CO$_2$ level ($d_2$) and the outdoor humidity ($d_4$), this noise is added directly to the weather data.
For the radiation ($d_1$) and the outdoor temperature ($d_3$), special care is required to ensure the perturbed weather profiles still follow a reasonable day/night cycle.
To achieve this, Brownian excursions \cite{vervaat1979relation}, stochastic processes that have constrained initial and final values, are used such that the Brownian noise only takes effect during the day cycle, and arrives at the relative night-time values at the start of the night cycle.
Figure~\ref{fig:disturbance_profiles} demonstrates the original, and five perturbed, weather disturbance profiles for the last two days of the growth cycle.
\begin{figure}
	\centering
	\input{media/tikz/disturbance_profiles}
	\caption{Last two days of weather profiles. The black profile is the original, and the red lines are five example perturbed profiles.}
	\label{fig:disturbance_profiles}
\end{figure}

\subsection{Comparison Approaches}\label{sec:comparison_approaches}
We compare our approach, referred to as MPC-RL, against the robust-sample based MPC controller from \cite{boersmaRobustSampleBasedModel2022}, and a model-free RL controller trained with DDPG \cite{morcegoReinforcementLearningModel2023b, wangDeepReinforcementLearning2020}.
Furthermore, we include two nominal MPC controllers for comparison.
One is ideal, with perfect (but unrealistic) model knowledge, while the other is a nominal MPC controller using an incorrect prediction model.
\begin{color}{black}All MPC controllers in the following use a horizon of $N=24$.
This value is selected to balance performance and computation time \cite{boersmaRobustSampleBasedModel2022, morcegoReinforcementLearningModel2023b}.\end{color}
In particular, the comparison approaches are:
\begin{itemize}
	\item \textbf{Ideal MPC controller (I-MPC)}: The ideal MPC controller is a standard MPC controller with knowledge of the true dynamics via the real values of $p$.
	\item \textbf{Nominal MPC controller (N-MPC)}: The nominal MPC controller is a standard MPC controller using an incorrect prediction model with parameters $\hat{p}\sim \mathcal{R}$.
	\item \textbf{Robust MPC controller (R-MPC-$n$) \cite{boersmaRobustSampleBasedModel2022}}: The robust sample-based MPC controller is essentially an implementation of a stochastic MPC controller based on the scenario approach \cite{schildbachScenarioApproachStochastic2014}.
	The parametric model uncertainty is addressed by sampling the probability distribution of model parameters, in this case the uniform distribution $\mathcal{R}$, and optimizing one state trajectory for each sample in the MPC optimization.
	In \cite{boersmaRobustSampleBasedModel2022}, 20 samples were used. 
	In our experiments we test a range of numbers of samples $n$, indicating the number of samples in the shorthand name, e.g., 5 samples is denoted R-MPC-5.
	\item \textbf{DDPG RL controller (DDPG) \cite{morcegoReinforcementLearningModel2023b, wangDeepReinforcementLearning2020}}: DDPG is a model-free off-policy RL algorithm that leverages DNNs as function approximators. Conversely to the proposed Q-learning algorithm, this method belongs to the policy gradient family, which optimizes the parametrization via estimates of the gradient of the policy w.r.t. $\theta$ \cite{lillicrap2015continuous}. The learning hyper-parameters are taken from \cite{morcegoReinforcementLearningModel2023b} and are reported in Table \ref{tab:ddpg_hyperparams}.
\end{itemize}
{\color{black}Note that both learning-based and non-learning-based controllers described above are deployed in simulations on the same greenhouse environment. Obviously, this environment employs the correct model to simulate the evolution of the real dynamics and to generate training data.}

\subsection{Results} \label{sec:results}
\begin{table}
    \caption{Hyper-parameters for the DDPG RL Controller.}
    \centering
    \begin{tabular}{lc}
        \toprule
        Parameter                   & Value     \\
        \hline 
        learning rate               & $10^{-5}$ \\
        gradient threshold          & 1 \\
        $L_2$ regularization factor & $10^{-5}$ \\
        experience buffer size      & $10^4$ \\
        experience mini-batch size  & 64 \\
        \bottomrule
    \end{tabular}
    \label{tab:ddpg_hyperparams}
\end{table}
We run 100 growth cycles to learn the parametrization of the MPC scheme \eqref{eq:MPC_parametrized}, with the initial parametrization in Table \ref{tab:initialization}.
{\color{black}We simulate online learning, where the data generated while interacting with the system is used to learn the policy. The learning is concluded after 100 episodes as, for this case study, this is sufficient to observe satisfactory convergence of the TD error and the learned MPC parametrization, alongside improvements to both the economic profit and constraint satisfaction.}
The hyper-parameters of the learning process are tuned as follows.
The discount factor is selected as $\gamma = 0.99$, while the learning rate is $\alpha=0.1$.
The parameters are updated once at the end of each growth cycle, with the maximum update  being constrained to 5\% of their current values, i.e., $\Delta \theta_\text{lb} = -0.05|\theta|$ and $\Delta \theta_\text{ub} = 0.05|\theta|$.
A replay buffer stores the observations from the last three cycles, and every update considers two cycles worth of observations, with all observations from the most recent cycle guaranteed to be used.
Three different parametrizations are learned for both RL-based approaches, where, for each, the random seed for generating the weather profiles is different.
The initial state for each cycle is $x(0) = \begin{pmatrix} 0.0035 & 0.001 & 15 &0.008 \end{pmatrix}^T$. 

Figure~\ref{fig:training} shows the episode-wise stage cost $L_\text{ep} = \sum_{k = 0}^{N_\text{s}} L\big(s(k), u(k)\big)$, the TD error $\delta_\text{ep} = 1/N_\text{s} \sum_{k=0}^{N_\text{s}} \delta(k)$, and the constraint violations $\Psi$ for each growth cycle during training.
It can be seen that the incurred stage cost and TD error are significantly reduced during training.
This is further represented in the constraint violations, which, starting from a large value, approach zero.
Figure~\ref{fig:outputs} shows the constrained outputs during the first and last growth cycles of training.
It can be seen that in the first growth cycle, the indoor CO$_2$ upper bound ($y_2$) is violated several times towards the end of the cycle, while the indoor humidity upper bound ($y_4$) is violated for practically the entire cycle.
After 100 cycles of learning, it can be seen that the indoor CO$_2$ level never violates the bounds, and the indoor humidity level exceeds the upper bound only a few times, quickly dropping back to allowable levels.
Figure~\ref{fig:parameters} shows the parameter evolution over the training for a subset of the learnable parameters; the four parameters whose values changed the most.
\begin{color}{black}Finally, in Appendix~\ref{ap:trajs} the control inputs during the first and last growth cycles of training are included for completeness.\end{color}
\begin{figure}
	\centering
	\input{media/tikz/training}
	\caption{Stage cost, TD error, and constraint violations over 100 growth cycles of training.}
	\label{fig:training}
\end{figure}
\begin{figure*}
	\centering
	\input{media/tikz/outputs_slimmer}
	\caption{Outputs over first and last growth cycle of training with upper (red lines) and lower (black lines) bounds.}
	\label{fig:outputs}
\end{figure*}
\begin{figure}
	\centering
	\input{media/tikz/parameters}
	\caption{Parameter evolution, for the 4 most changed parameters, over 100 growth cycles of training. Note that model parameters $\theta_p$ have been normalized with respect to their true values.}
	\label{fig:parameters}
\end{figure}

\begin{color}{black}In an evaluation phase, i.e., with the policy fixed, we compare our final trained RL-based MPC controller against the approaches outlined in Section~\ref{sec:comparison_approaches}.\end{color}
Each approach is evaluated over 100 growth cycles, with Figure~\ref{fig:comparison} demonstrating the results with respect to the performance indicators given in Section~\ref{sec:problem_formulation}.
The episode-wise stage cost is included for interest, especially for comparison between the two RL-based approaches, that during training learn policies to minimize this cost, as in \eqref{eq:RL_optimal_pol}. 
\begin{color}{black}For completeness,  the state, output, and input trajectories for the final 10 days of an evaluation growth cycle are included in Appendix~\ref{ap:trajs}.\end{color}

The constraint violations $\Psi$ show that our approach has reduced the constraint violations the most, approaching the level of the ideal MPC controller. 
The robust MPC controller does improve the constraint violations over the nominal MPC; however, they remain at an unsatisfactory level.
Finally, the DDPG-based controller improves the violations over both the nominal and robust MPC controllers, but does not reach that of our approach.

The crop yield $y_1(N_\text{s})$ shows that the controllers with very high constraint violations, i.e., the nominal MPC controller and robust MPC controller with 5 samples, generate a very high crop yield. 
This is in line with observations in the literature \cite{xuAdaptiveTwoTimeScale2018}, and is due to the unrealistic growth that occurs in the model when variables such as CO$_2$ levels and humidity are dangerously high.
Notably, for higher numbers of samples, the robust MPC controllers significantly reduce the crop yield due to conservatism.
Our approach has reduced the drop in crop yield, even while vastly improving the constraint violations.
The DDPG controller gives superior crop yield even with respect to the ideal MPC controller.
This is again in line with observations in the literature \cite{morcegoReinforcementLearningModel2023b}, and is due to the DDPG approach's tendency to optimize the crop yield over efficiency and constraint violations.

The economic profit indicator $P$ shows similar trends to the crop yield for all MPC-based controllers.
\begin{color}{black}Again, we highlight that the high economic profit of the nominal MPC controller, and the robust MPC controller with 5 samples, is due to the unrealistic growth occurring in the model when the output constraints are significantly violated.\end{color}
A clear difference from the crop yield trend is for the DDPG controller, which now under-performs in comparison to both the ideal MPC controller and our approach.
This demonstrates how the DDPG controller maximizes the yield in an inefficient way with respect to the control inputs, resulting in an overall reduced profit.

Finally,  with respect to the stage cost $L_\text{ep}$, our approach outperforms the DDPG controller.
This is notable as, during training, both RL-based approaches attempt to solve an optimization problem that minimizes $L_\text{ep}$, i.e., the performance \eqref{eq:RL_performance}.
\begin{figure}
	\centering
	\input{media/tikz/comparison}
	\caption{Comparison of all approaches over 100 growth cycles. The learning-based strategies are tested after convergence. Bars show the mean value, with error bars showing a standard deviation.}
	\label{fig:comparison}
\end{figure}

Table \ref{tab:solve_times} shows the average computation time needed per time step for each of the approaches.
Naturally, the DDPG approach is the fastest as no optimization problem is solved when computing its actions.
\begin{color}{black}Comparing the MPC-based approaches, our approach requires similar computation times to those of the I-MPC and N-MPC approaches during evaluation, and is only slightly slower during training.\end{color}
In contrast, the R-MPC approaches introduce additional optimization variables with each additional sample, and the required computation time increases as the samples increase.
\begin{table*}
	\caption{Average computation time required by each approach.}
	\begin{center}
		\begin{tabular}{l|cccccccc}
			\toprule 	
			Approach & I & N & R5 & R10 & R20 & \begin{color}{black}\makecell{MPC-RL \\ training}\end{color} & \begin{color}{black}\makecell{MPC-RL \\ evaluation}\end{color} & DDPG \\
			\midrule		
			Time (s) & 0.0378 & 0.0390 & 0.229 & 0.570 & 1.234 & \begin{color}{black}0.0793\end{color} & \begin{color}{black}0.0430\end{color} & 0.000155 \\
			\bottomrule
		\end{tabular}
		\label{tab:solve_times}
	\end{center}\end {table*}
\section{Conclusions}\label{sec:conclusions}

In this work, we propose an RL-based MPC controller for greenhouse climate control.
At the core of the approach is a parametrized MPC scheme that can serve as policy provider and value-function approximator for an RL task.
The greenhouse climate control problem has been formulated as an RL task, such that the MPC scheme can learn a parametrization online using closed-loop data.
Second-order Q-learning has been proposed as the RL algorithm for learning the parametrization.
In simulations, the proposed approach has been shown to learn an MPC scheme that significantly reduces the violations of output constraints in closed-loop operation.
The final learned controller has then been compared against state-of-the-art MPC- and RL-based controllers from the literature, showing the best performance in terms of constraint violations and efficient crop growth. 

Future work directions include the application of this methodology to alternative greenhouse models, and experimental validation in real-world tests.
Additionally, alternative learning algorithms, such as policy gradient approaches, could be explored to learn the parametrization of the MPC scheme.

\section{Funding}
This paper is part of a project that has received funding from the European Research Council (ERC) under the European Union’s Horizon 2020 research and innovation programme (Grant agreement No. 101018826
- CLariNet).

\bibliographystyle{plain}  
\bibliography{references}  
\appendix
\section*{Appendix}\label{sec:appendix}

\section{Extra Input, Output, and State Trajectories} \label{ap:trajs}
\begin{color}{black}Figure~\ref{fig:inputs_train} shows the control inputs of our approach during the first and last episodes of training. 
The trajectories are clearly different, as the approach has learned a policy that optimizes for closed-loop performance.
\begin{figure}
	\centering
	\input{media/tikz/inputs}
    \captionsetup{labelfont={color=black}, textfont={color=black}}
	\caption{Control inputs for the first and last growth cycle of training with upper (red lines) and lower (black lines) bounds.}
	\label{fig:inputs_train}
\end{figure}

Figures \ref{fig:outputs_comp}, \ref{fig:states_comp}, and \ref{fig:inputs_comp} show the output, state, and input trajectories, respectively, for the final 10 days of a growth cycle during evaluation of the compared controllers.
Notably, the additional constraint violations in the outputs, introduced by the comparison controllers, are evident. 
Further, the control inputs demonstrate the inefficiency of the DDPG controller's actuation.
\begin{figure}
	\centering
    \input{media/tikz/outputs_comparison}
    \captionsetup{labelfont={color=black}, textfont={color=black}}
	\caption{Output trajectories for the last 10 days of a growth cycle during evaluation, with upper (red lines) and lower (black lines) bounds.}
	\label{fig:outputs_comp}
\end{figure}
\begin{figure}
	\centering
	\input{media/tikz/states_comparison}
    \captionsetup{labelfont={color=black}, textfont={color=black}}
	\caption{State trajectories for the last 10 days of a growth cycle during evaluation.}
	\label{fig:states_comp}
\end{figure}
\begin{figure}
	\centering
	\input{media/tikz/inputs_comparison}
    \captionsetup{labelfont={color=black}, textfont={color=black}}
	\caption{Input trajectories for the last 10 days of a growth cycle during evaluation, with upper (red lines) and lower (black lines) bounds.}
	\label{fig:inputs_comp}
\end{figure}
\end{color}

\section{Greenhouse Model}
The continuous-time lettuce growing greenhouse model is described by the following set of equations \cite{vanhentenGreenhouseClimateManagement1994}
\begin{equation*}
	\begin{aligned}
		\dot{x}_1(t) ={}& p_{1} \phi_{\text{phot,c}}(t) -  p_{2} x_1(t) 2^{x_3(t)/10-5/2}, \\
		\dot{x}_2(t) ={}& \frac{1}{p_{9}}   \Big( -\phi_{\text{phot,c}}(t) + p_{10} x_1(t)  2^{x_3(t)/10-5/2}  \\ 
		& + 10^{-6} u_1(t) - \phi_{\text{vent,c}}(t) \Big), \\
		\dot{x}_3(t) ={}& \frac{1}{p_{16}} \Big( u_3(t) - \bigl(10^{-3} p_{17} u_2(t) +p_{18}\bigr), \\
		& \cdot \bigl(x_3(t)-d_3(t)\bigr) + p_{19} d_1(t) \Big), \\
        \dot{x}_4(t) ={}& \frac{1}{p_{20}} \big( \phi_{\text{transp,h}}(t) - \phi_{\text{vent,h}}(t) \big) , \\
		y_{1}(t) ={}& 10^3 x_1(t), \\
		y_{2}(t) ={}& 10^3 x_2(t) \frac{p_{12} \bigl(x_3(t)+p_{13}\bigr)}{p_{14}p_{15}}, \\
		y_{3}(t) ={}& x_3(t), \\
		y_{4}(t) ={}& \frac{10^2}{11} x_4(t) \frac{p_{12} \bigl(x_3(t) + p_{13}\bigr)}{\exp{\left(\frac{p_{27}x_3(t)}{x_3(t) + p_{28}}\right)}}.
	\end{aligned}
\end{equation*}
where $t \geq 0$ is continuous time.
Further, define the following functions:
\begin{equation*}
	\begin{aligned}
		\phi_{\text{phot,c}}(t) ={}& \frac{1}{\varphi(t)} \Big(1-\text{exp} \big(-p_{3}x_1(t) \big) \Big) \\
		& \cdot\Big(p_{4}d_1(t)\big(-p_{5}x_3(t)^2+p_{6} x_3(t) - p_{7} \big) \\
		& \cdot\big( x_2(t)- p_{8}\big)\Big), \\
		\varphi(t) ={}& p_{4}d_1(t)+\big(-p_{5}x_3(t)^2+p_{6} x_3(t) - p_{7} \big)\\
		& \cdot \big( x_2(t)- p_{8}\big), \\
		\phi_{\text{vent,c}}(t) ={}& \big(u_2(t)10^{-3}+p_{11}\big)\big(x_2(t)-d_2(t)\big),\\
		\phi_{\text{transp,h}}(t) ={}& p_{21} \Big(1-\exp{\big(-p_{3}x_1(t) \big)} \Big) \\ 
		& \hspace{-1.3cm} \cdot \Bigg(\frac{p_{22}}{p_{23}(x_3(t)+p_{24})} \exp{\left( \frac{p_{25} x_3(t)}{x_3(t) + p_{26}} \right)} - x_4(t) \Bigg), \\
		\phi_{\text{vent,h}}(t) ={}& \big(u_2(t)10^{-3}+p_{11}\big)\big(x_4(t)-d_4(t)\big),		
	\end{aligned}
\end{equation*}
where $\phi_{\text{phot,c}}(t), \phi_{\text{vent,c}}(t),$ $\phi_{\text{transp,h}}(t)$ and $\phi_{\text{vent,h}}(t)$ are the gross canopy photosynthesis rate, mass exchange of CO$_2$ through the vents, canopy transpiration and mass exchange of H$_2$O through the vents, respectively.

To generate the discrete-time prediction mode \eqref{eq:true_model}, discretization is performed using the explicit fourth order Runge-Kutta method with a time step of 15 minutes, as in \cite{boersmaRobustSampleBasedModel2022}.
Finally, the values for the model parameters $p$ are given in Table~\ref{tab:model_parameters}. 
\begin{table*}
	\centering
	\caption{Model parameters.}
    {\color{black}
	\begin{tabularx}{\textwidth}{lclX} 
            \toprule
            Symbol   & Value & Unit & Explanation from \cite{vanhentenGreenhouseClimateManagement1994,vanhenten2003sensitivity}  \\
			\midrule
			$p_{1}$  & $0.544$              & -                                                   & Yield factor; aggregation of $c_\alpha$ ($\text{CO}_2$-glucose stoichiometric conversion factor) and $c_\beta$ (yield of carbohydrates conversion to structural material)                                               \\  
			$p_{2}$  & $2.65 \cdot 10^{-7}$ & \unit{\per\second}                                  & Aggregation of $c_\text{resp,s}$ and $c_\text{resp,r}$ (maintenance respiration rates for shoot and root at \qty{25}{\celsius}) weighted by $c_\beta$ and $c_\tau$ (ratio of root dry weight to total crop dry weight)  \\  
			$p_{3}$  & $53$                 & \unit{\square\meter\per\kilogram}                   & Effective canopy surface; aggregation of $c_\text{lar,d}$ (shoot leaf area ratio) with $c_\text{k}$ (canopy extinction coefficient) and $c_\tau$                                                                        \\  
			$p_{4}$  & $3.55\cdot 10^{-9}$  & \unit{\kilogram\per\joule}                          & Aggregation of $c_\text{par}$ (photosynthetically activate radiation to total solar radiation ratio) and $c_\text{rad,rf}$ (roof solar radiation transmission coefficient)                                              \\  
			$p_{5}$  & $5.11\cdot 10^{-6}$  & \unit{\meter\per\second\per\square\celsius}         & Second-order term in polynomial approximation of temperature effect on $\text{CO}_2$ leaf diffusion                                                                                                                     \\  
			$p_{6}$  & $2.3\cdot 10^{-4}$   & \unit{\meter\per\second\per\celsius}                & First-order term in polynomial approximation of temperature effect on $\text{CO}_2$ leaf diffusion                                                                                                                      \\  
			$p_{7}$  & $6.29\cdot 10^{-4}$  & \unit{\meter\per\second}                            & Zeroth-order term in polynomial approximation of temperature effect on $\text{CO}_2$ leaf diffusion                                                                                                                     \\  
			$p_{8}$  & $5.2\cdot 10^{-5}$   & \unit{\kilogram\per\cubic\meter}                    & $\text{CO}_2$ compensation point at \qty{25}{\celsius}                                                                                                                                                                  \\  
			$p_{9}$  & $4.1$                & \unit{\meter}                                       & Volumetric capacity of greenhouse air for $\text{CO}_2$                                                                                                                                                                 \\  
			$p_{10}$ & $4.87\cdot 10^{-7}$  & \unit{\per\second}                                  & Aggregation of $c_\text{resp,s}$ and $c_\text{resp,r}$ weighted by $c_\alpha$ and $c_\tau$                                                                                                                              \\  
			$p_{11}$ & $7.5\cdot 10^{-6}$   & \unit{\meter\per\second}                            & Leakage air exchange through greenhouse cover                                                                                                                                                                           \\  
			$p_{12}$ & $8.314$              & \unit{\joule\per\kelvin\per\mol}                    & Gas constant                                                                                                                                                                                                            \\  
			$p_{13}$ & $273.15$             & \unit{\kelvin}                                      & Absolute temperature                                                                                                                                                                                                    \\  
			$p_{14}$ & $101325$             & \unit{\pascal}                                      & Sea level standard atmospheric pressure                                                                                                                                                                                 \\  
            $p_{15}$ & $0.044$              & \unit{\kilogram\per\mol}                            & $\text{CO}_2$ molar mass                                                                                                                                                                                                \\  
            $p_{16}$ & $3\cdot 10^{4}$      & \unit{\joule\per\square\meter\per\celsius}          & Heat capacity of greenhouse air                                                                                                                                                                                         \\  
            $p_{17}$ & $1290$               & \unit{\joule\per\cubic\meter\per\celsius}           & Heat capacity per volume unit of greenhouse air                                                                                                                                                                         \\  
            $p_{18}$ & $6.1$                & \unit{\watt\per\square\meter\per\celsius}           & Heat transfer coefficient through greenhouse cover                                                                                                                                                                      \\  
            $p_{19}$ & $0.2$                & -                                                   & Sun heat load coefficient accounting for roof transmission, solar radiation interception by structural components, and conversion from absorbed solar energy to sensible heat load by canopy                            \\  
            $p_{20}$ & $4.1$                & \unit{\meter}                                       & Volumetric capacity of greenhouse air for humidity                                                                                                                                                                      \\  
            $p_{21}$ & $0.0036$             & \unit{\meter\per\second}                            & Canopy transpiration mass transfer coefficient                                                                                                                                                                          \\  
            $p_{22}$ & $9348$               & \unit{\joule\kilogram\per\cubic\meter\per\kilo\mol} & Aggregation of $c_{\text{H}_2 \text{O}}$ (water molar mass), $c_{\text{v},0}$ (calibration parameter) and $c_{\text{v},1}$ (saturation water vapor pressure parameter for canopy transpiration)                         \\  
            $p_{23}$ & $8314$               & \unit{\joule\per\kelvin\per\kilo\mol}               & Gas constant                                                                                                                                                                                                            \\  
            $p_{24}$ & $273.15$             & \unit{\kelvin}                                      & Absolute temperature                                                                                                                                                                                                    \\  
            $p_{25}$ & $17.4$               & -                                                   & Saturation water vapor pressure parameter for canopy transpiration                                                                                                                                                      \\  
            $p_{26}$ & $239$                & \unit{\celsius}                                     & Saturation water vapor pressure parameter for canopy transpiration                                                                                                                                                      \\  
            $p_{27}$ & $17.269$             & -                                                   & Saturation water vapor pressure parameter for humidity                                                                                                                                                                  \\  
            $p_{28}$ & $238.3$              & \unit{\celsius}                                     & Saturation water vapor pressure parameter for humidity                                                                                                                                                                  \\  
            \bottomrule
	\end{tabularx}
    \label{tab:model_parameters}
    }
\end{table*}

\clearpage
\appendix

\end{document}